\documentclass[10pt,conference]{IEEEtran}

\usepackage[left=0.625in,right=0.625in,top=0.7in,bottom=1in]{geometry}
\usepackage{amsmath, graphics,amssymb,epsfig,subfigure,color,amsthm,cite}
\usepackage{array}
\usepackage{multirow}
\usepackage{enumerate}
\usepackage{algorithmic}
\usepackage{algorithm}
\usepackage{bm}
\usepackage{algorithmic}
\usepackage[caption=false]{subfig}
\newtheorem{thm}{Theorem}

\begin{document}
%\ifonecol
%    \title{\LARGE{Robust MLSD for Wideband SIMO Systems with One-bit ADCs: Reinforcement Learning Approach}}
%    \vspace{-2mm}
%\else
	\title{Data-Aided Channel Estimator for MIMO Systems via Reinforcement Learning}
%\fi
%
%\author{Yo-Seb Jeon, Minji So, and Namyoon Lee
%\thanks{Y.-S. Jeon, M. So, and N. Lee are with the Department of Electrical Engineering, POSTECH, Pohang, Gyeongbuk, 37673 Korea (e-mail: yoseb.jeon@postech.ac.kr, mjso@postech.ac.kr, nylee@postech.ac.kr).}}

\author{\IEEEauthorblockN{Yo-Seb Jeon\IEEEauthorrefmark{1}, Jun Li\IEEEauthorrefmark{2}, Nima Tavangaran\IEEEauthorrefmark{3}, and H. Vincent Poor\IEEEauthorrefmark{3}}
	\IEEEauthorblockA{\IEEEauthorrefmark{1}Department of Electrical Engineering, POSTECH, Gyeongbuk 37673, South Korea}
	\IEEEauthorblockA{\IEEEauthorrefmark{2}Department of Electronic and Optical Engineering, Nanjing University of Science and Technology, Nanjing 210094, China}
	\IEEEauthorblockA{\IEEEauthorrefmark{3}Department of Electrical Engineering, Princeton University, Princeton, NJ 08544\\
		Email: yoseb.jeon@postech.ac.kr, jun.li@njust.edu.cn, nimat@princeton.edu, poor@princeton.edu}}

\vspace{-4mm}

\maketitle

\newcommand{\argmax}{\operatornamewithlimits{argmax}}
\newcommand{\argmin}{\operatornamewithlimits{argmin}}
\makeatletter
\def\blfootnote{\gdef\@thefnmark{}\@footnotetext}
\makeatother
\vspace{-14mm}

\begin{abstract}
This paper presents a data-aided channel estimator that reduces the channel estimation error of the conventional linear minimum-mean-squared-error (LMMSE) method for multiple-input multiple-output communication systems. The basic idea is to selectively exploit detected symbol vectors obtained from data detection as additional pilot signals. To optimize the selection of the detected symbol vectors, a Markov decision process (MDP) is defined which finds the best selection to minimize the mean-squared-error (MSE) of the channel estimate. Then a reinforcement learning algorithm is developed to solve this MDP in a computationally efficient manner. Simulation results demonstrate that the presented channel estimator significantly reduces the MSE of the channel estimate and therefore improves the block error rate of the system, compared to the conventional LMMSE method. 
\end{abstract}

%\begin{IEEEkeywords}
%Maximum-likelihood detection (MLD), analog-to-digital converter (ADC), one-bit ADC, channel estimation, classification.
%\end{IEEEkeywords}

%%%%%% Acknowledgement
\blfootnote{This work was supported in part by Basic Science Research Program through the National Research Foundation of Korea (NRF) funded by the Ministry of Education (2019R1A6A3A12032717), in part by the German Research Foundation (DFG) under Grant TA 1431$/$1-1, and in part by the  U.S. National Science Foundation under Grants CCF-0939370 and CCF-1513915.} 
%This work was supported by the  U.S. National Science Foundation under Grants CCF-0939370 and CCF-1513915.}

\section{Introduction}
%%%%%% MIMO system, and importance of channel information
Multiple-input multiple-output (MIMO) communication is one of the core technologies in modern wireless standards as it significantly improves both the capacity and the reliability of wireless systems by providing spatial multiplexing and diversity gains \cite{Foschini:96,Telatar:99,Zheng:03}. A key requirement to enjoy these benefits is accurate channel state information (CSI) at both transmitter and receiver. For example, the capacity of MIMO communication systems increases linearly with the number of either transmit or receive antennas under the premise that perfect CSI is available at both the transmitter and the receiver \cite{Foschini:96,Telatar:99}. Various techniques to enable the accurate CSI have been extensively developed for the MIMO systems to fully achieve their potential gains \cite{Biguesh:06,Ozdemir:07,Dowler:03,Zhao:08,Ma:14,Park:15}.

%%%%%% Pilot-aided method -> Data-aided method 
Pilot-aided channel estimation is one of the most popular and widely adopted techniques to obtain the CSI at the receiver (CSIR) \cite{Biguesh:06,Ozdemir:07}. The fundamental idea is to send pilot signals that are priorly known at the receiver and then to estimate the CSI based on the information of the pilot signals and the corresponding received signals. A representative example of this technique is the least-squares (LS) channel estimator that minimizes the sum of squared errors in the channel estimate \cite{Biguesh:06,Ozdemir:07}. Another example is the linear minimum-mean-squared-error (LMMSE) channel estimator which is a linear estimator that minimizes the mean-squared-error (MSE) of the channel estimate based on the statistical model \cite{Biguesh:06,Ozdemir:07}. 
%By exploiting the second-order statistic of the channel, the LMMSE channel estimator performs better than the LS channel estimator. 
The accuracy of the CSIR obtained from these pilot-aided estimators heavily depends on a pilot signal length allowed in the system. Unfortunately, in practice, the pilot signal length is very limited because the transmission of the pilot signals also consumes radio resources; thereby, the use of the pilot-aided channel estimator with a limited pilot length suffers from inevitable channel estimation errors. 
%Unfortunately, the pilot signal length is limited in practical systems because the amount of radio resources available for data transmission reduces with the pilot length. 

Data-aided channel estimation has been proposed to overcome the limitation of the pilot-aided method \cite{Dowler:03,Zhao:08,Ma:14,Park:15}. In this technique, detected symbols obtained from data detection at the receiver are exploited for updating the channel estimate, so the accuracy of the CSIR can be improved without increasing the pilot signal length. A non-iterative data-aided channel estimator was developed in \cite{Dowler:03} which exploits data symbols reconstructed at the receiver, but such non-iterative approach is vulnerable to error propagation caused by data detection errors. To resolve this problem, an iterative approach has been adopted in most existing data-aided channel estimators \cite{Zhao:08,Ma:14,Park:15}. In this approach, the channel estimation and the data detection are performed iteratively to improve the accuracy of both the channel estimate and the detected symbols. In \cite{Zhao:08}, an iterative turbo channel estimation technique was developed in which soft-decision symbols are utilized as pilot signals at each iteration. A similar iterative approach was proposed in \cite{Park:15} where the soft-decision symbols are selectively utilized as the pilot signals by using an MSE-based selection method. The common limitation of these iterative methods is that they increase not only the computational complexity at the receiver but also communication latency.   
%To the best knowledge of the authors, none of the existing work develops a data detection method that is robust to temporal channel variations in MIMO systems with one-bit ADCs.    
%

%the common limitation of the methods in \cite{Jeon:TVT:18,Nguyen:18,Choi:18,Hong:19,Jeon:arXiv:19} is that they are vulnerable to temporal channel variations caused by fast changes in the communication environment or by a receiver with high mobility.  
%%shown to be effective when a channel coherence time is sufficiently large. Unfortunately, this condition may not hold when channel variations over time are high due to the mobility of the receiver or a fast change in a communication environment. 
%To the best knowledge of the authors, none of the existing work develops a data detection method that is robust to temporal channel variations in MIMO systems with one-bit ADCs.    
%

%%%%%% Proposed robust Q-Viterbi algoprithm
%%%%%%
In this paper, we propose a data-aided channel estimator that reduces the channel estimation error of the conventional LMMSE channel estimator for MIMO systems. The basic idea is to selectively exploit detected symbols obtained from data detection as additional pilot signals. Although this idea is similar to that of the existing data-aided channel estimators, 
%We first characterize the first-order and second-order statistics of the likelihood functions at each time slot based on the statistics at a previous time slot. We then update these statistics by exploiting the input-output samples, each contains the information of the likelihood function at ech time slot. 
the key difference is that a reinforcement learning approach is adopted to optimize the selection of the detected symbols, inspired by the work in \cite{Jeon:arXiv:19}. 
To this end, we first define a Markov decision process (MDP) which finds the best selection to minimize the MSE of the channel estimate. We then derive a closed-form policy to solve this MDP based on a reinforcement learning approach. The prominent feature of the proposed method is that it mitigates the error propagation effect in the data-aided channel estimation even without the aid of an iterative approach. 
%The prominent feature of the proposed method is that the error propagation effect is captured when optimizing the detected-symbol selection even without using an iterative approach.
% it optimizes the decision on whether to use each detected symbol by taking into account the error propagation effect, but in a non-iteartive way. 
Simulation results show that the proposed method significantly reduces the MSE of the channel estimate at the receiver and therefore improves the block error rate of the system, compared to the conventional LMMSE method.

%\subsubsection*{Notation}
%Upper-case and lower-case boldface letters denote matrices and column vectors, respectively.
%$\mathbb{E}[\cdot]$ is the statistical expectation,
%    $\mathbb{P}(\cdot)$ is the probability,
%    $(\cdot)^\top$ is the transpose,
%    $|\cdot|$ is the absolute value,
%    $\text{Re}(\cdot)$ is the real part,
%    $\text{Im}(\cdot)$ is the imaginary part,
%    and $\lfloor\cdot\rfloor$ is the floor function.
%
%\begin{figure}
%	\centering
%	\includegraphics[width=3.5in]{Fig_Frame.eps} \vspace{-0.3cm}\caption{An illustration of a transmission frame considered in this paper, which consists of one pilot frame and $N_{\rm d}$ data frames.} \vspace{-3mm} \label{fig:Frame}
%\end{figure}

%%%%%%%%%%%%%%%%%%%%%%%%%%%%%%%%%%%%%%%%%%% %%%%%%%%%%%%%%%%%%%%%%%%%%%%%%%%%%%%%%%
\section{System Model and Preliminaries}
In this section, we introduce the system model considered in this work. We then present the LMMSE channel estimator and the maximum-a-posteriori-probability (MAP) data detector for the considered system.

\subsection{System Model}\label{Sec:System}
We consider a coded MIMO communication system in which a transmitter equipped with ${N}_{\rm tx}$ antennas communicates with a receiver equipped with $N_{\rm rx}$ antennas. We model the wireless channel of the considered system as a frequency-flat Rayleigh-fading channel denoted by ${\bf H}=[{\bf h}_1,\cdots,{\bf h}_{N_{\rm rx}}]\in\mathbb{C}^{N_{\rm tx}\times N_{\rm rx}}$,
%\begin{align}
%	{\bf H}=[{\bf h}_1,\cdots,{\bf h}_{N_{\rm rx}}]\in\mathbb{C}^{N_{\rm tx}\times N_{\rm rx}},
%\end{align}
where ${\bf h}_{r}\sim\mathcal{CN}({\bf 0}_{N_{\rm tx}},{\bf I}_{N_{\rm tx}})$ is the wireless channel between the transmitter and the $r$-th receive antenna. We assume a block-fading channel in which the elements of ${\bf H}$ keep constant during a transmission frame. 

%As illustrated in Fig.~\ref{fig:Block}, 
We consider a transmission frame that consists of one pilot block with length $T_{\rm p}$ and $N_{\rm B}$ data blocks each with length $T_{\rm d}$. A set of time slot indices associated with the pilot block and the $b$-th data block is denoted as $\mathcal{N}_{\rm p}=\{1,\ldots,T_{\rm p}\}$ and $\mathcal{N}_b=\{T_{\rm p}+(b-1)T_{\rm d}+1,\ldots,T_{\rm p}+bT_{\rm d}\}$,   
%\begin{align}\label{eq:N_b}
%	\mathcal{N}_b=\{T_{\rm p}+(b-1)T_{\rm d}+1,\ldots,T_{\rm p}+bT_{\rm d}\},
%\end{align}
respectively, for $b \in\{1,\ldots,N_{\rm B}\}$. Let ${\bf x}^{\rm p}[n]\in\mathbb{C}^{N_{\rm tx}}$ be the pilot signal sent at time slot $n$ such that $\mathbb{E}[\|{\bf x}^{\rm p}[n]\|^2]=N_{\rm tx}$. Then the received signal associated with ${\bf x}^{\rm p}[n]$ is given by 
\begin{align}\label{eq:Y_p}
	{\bf y}^{\rm p}[n] =\big[y_1^{\rm p}[n],\cdots, y_{N_{\rm rx}}^{\rm p}[n]\big]^{\top} 
	= {\bf H}^{\sf H}{\bf x}^{\rm p}[n] + {\bf z}[n],
\end{align}
for $n\in\mathcal{N}_{\rm p}$, where ${\bf z}[n]\sim \mathcal{CN}({\bf 0}_{N_{\rm rx}},{\sigma^2}{\bf I}_{N_{\rm rx}})$ is a circularly symmetric complex Gaussian noise vector at time slot $n$. 
%Then the receiver uses the prior knowledge of the pilot signals to estimate the wireless channel ${\bf H}$. 
For the transmission of each data block, the transmitter generates $T_{\rm d}$ data symbol vectors by applying 1) CRC appending, 2) channel encoding, and 3) symbol mapping to information bits. Let ${\bf x}[n]\in \mathcal{X}^{N_{\rm tx}}$ be the data symbol vector sent at time slot $n\in\mathcal{N}_b$ for $b\in\{1,\ldots,N_{\rm B}\}$, where $\mathcal{X}$ is a constellation set such that $\mathbb{E}[\|{\bf x}[n]\|^2]=N_{\rm tx}$. Then the received signal associated with ${\bf x}[n]$ is given by
\begin{align}\label{eq:y_n}
	{\bf y}[n] =\big[y_1[n],\cdots, y_{N_{\rm rx}}[n]\big]^{\top} 
	= {\bf H}^{\sf H}{\bf x}[n] + {\bf z}[n],
\end{align}
for $n\in\mathcal{N}_b$ and $b\in\{1,\ldots,N_{\rm B}\}$.
%, where ${\bf z}[n]\sim \mathcal{CN}({\bf 0}_{N_{\rm rx}},{\sigma^2}{\bf I}_{N_{\rm rx}})$ is a circularly symmetric complex Gaussian noise vector at time slot $n$. 

%\begin{figure*}
%	\centering 
%	{\epsfig{file=Figures/Fig_Block.eps, width=12cm}}
%	\caption{A transmission frame that consists of one pilot block with length $N_{\rm p}$ and $D$ data blocks each with length $N_{\rm d}$, where a guard interval with length $L-1$ is appended at the end of every block.} \vspace{-3mm}
%	\label{fig:Block}
%\end{figure*}

\subsection{LMMSE Channel Estimator}\label{Sec:CE}
The LMMSE channel estimator is a linear estimator that minimizes the MSE of the channel estimate, which has been widely adopted in wireless communication systems due to its fair performance with a low computational complexity \cite{Biguesh:06,Ozdemir:07}. 
%This channel estimator has been widely adopted in wireless communication systems as it provides a fair estimation performance with a low computational complexity. 
%In the considered system, the received signal at the $r$-th receive antenna associated with the pilot block is represented as 
%\begin{align}\label{eq:y_rp}
%	{\bf y}_r^{\rm p} &= \big[ y_r[1],\cdots,y_r[T_{\rm p}] \big] = {\bf h}_r^{\sf H}{\bf X}^{\rm p} + {\bf z}_r,
%\end{align}
%for $r\in\{1,\ldots,N_{\rm rx}\}$, where ${\bf X}^{\rm p} = \big[ {\bf x}^{\rm p}[1],\cdots,{\bf x}^{\rm p}[T_{\rm p}] \big]$,
%%\begin{align}\label{eq:X_p}
%%	{\bf X}^{\rm p} = \big[ {\bf x}^{\rm p}[1],\cdots,{\bf x}^{\rm p}[T_{\rm p}] \big],
%%\end{align}
%${\bf z}_r=\big[ z_r[1],\cdots,z_r[T_{\rm p}] \big]$, and $z_r[n]$ is the $r$-th element of ${\bf z}[n]$ for $r\in\{1,\ldots,N_{\rm rx}\}$.
%From \eqref{eq:y_rp}, the LMMSE filter is given by 
From \eqref{eq:Y_p}, the LMMSE filter of the considered system is given by  
\begin{align}\label{eq:W_LMMSE}
	{\bf W}_{\rm LMMSE} 
	&= \argmin_{{\bf W}\in\mathbb{C}^{N_{\rm tx}\times T_{\rm p}}} \!\mathbb{E} \Big[ \big\|{\bf W}({\bf y}_r^{\rm p})^{\sf H}  - {\bf h}_r\big\|^2\Big]
	\nonumber \\
	&= \big({\bf X}^{\rm p} ({\bf X}^{\rm p})^{\sf H} + \sigma^2 {\bf I}_{N_{\rm tx}}\big)^{-1}{\bf X}^{\rm p},
\end{align}
where ${\bf y}_r^{\rm p} = \big[ y_r^{\rm p}[1],\cdots,y_r^{\rm p}[T_{\rm p}] \big]$, ${\bf X}^{\rm p} = \big[ {\bf x}^{\rm p}[1],\cdots,{\bf x}^{\rm p}[T_{\rm p}] \big]$, and the expectation is taken with respect to the channel and the noise distributions. Consequently, the channel estimate obtained from the LMMSE channel estimator is computed as
\begin{align}\label{eq:h_hat_conv}
	\hat{\bf h}_{r} = \big({\bf X}^{\rm p} ({\bf X}^{\rm p})^{\sf H} + \sigma^2 {\bf I}_{N_{\rm tx}}\big)^{-1}{\bf X}^{\rm p} ({\bf y}_{r}^{\rm p})^{\sf H},
\end{align}
for $r\in\{1,\ldots,N_{\rm rx}\}$.

\subsection{Maximum-A-Posteriori-Probability (MAP) Data Detector}\label{Sec:MAP}
%The primary goal of the data detector for the considered system is to produce accurate log-likelihood ratios (LLRs), so that information bits sent at the transmitter are reconstructed by the channel decoder based on these LLRs. In this work, we focus on the optimal MAP data detector that produces the exact LLRs by computing a-posteriori-probabilities (APPs) for the given received signals. The major reason for this choice is to demonstrate the best performance that can be achieved by the proposed channel estimator which exploits the APPs obtained from the data detector.   
In this work, we focus on the optimal MAP data detector that computes a-posteriori-probabilities (APPs) for the given received signals. The major reason for this choice is to demonstrate the best performance that can be achieved by the proposed channel estimator which exploits the APPs obtained from the data detection.   
%In the optimal MAP data detector, the APPs are computed based on the channel information at the receiver. 

Let ${\bf x}_k$ be the $k$-th possible symbol vector in $\mathcal{X}^{N_{\rm tx}}$ for $k \in\mathcal{K}=\{1,\ldots,K\}$ where $K=|\mathcal{X}|^{N_{\rm tx}}$. The APP of the event $\{{\bf x}[n]={\bf x}_k\}$ for the given received signal ${\bf y}[n]$ is expressed as
\begin{align}\label{eq:APP_def}
	\theta_k[n] 
	&= \mathbb{P}\big[{\bf x}[n]\!=\!{\bf x}_k\big|{\bf y}[n]\big]  \nonumber \\
	&=  \frac{\mathbb{P}\big[{\bf y}[n]\big|{\bf x}[n]\!=\!{\bf x}_k\big]  \mathbb{P}\big[{\bf x}[n]\!=\!{\bf x}_k\big] }
		{\sum_{j\in\mathcal{K}} \mathbb{P}\big[{\bf y}[n]\big|{\bf x}[n]\!=\!{\bf x}_j\big] \mathbb{P}\big[{\bf x}[n]\!=\!{\bf x}_j\big] },
\end{align}
where $\mathbb{P}\big[{\bf y}[n]\big|{\bf x}[n]\!=\!{\bf x}_k\big]$ is the likelihood function that represents the probability of receiving ${\bf y}[n]$ for the given event $\{{\bf x}[n]={\bf x}_k\}$. The likelihood function $\mathbb{P}\big[{\bf y}[n]\big|{\bf x}[n]\!=\!{\bf x}_k\big]$ in the considered system is computed as
\begin{align}\label{eq:LF_def}
	\!\!\mathbb{P}\big[{\bf y}[n]\big|{\bf x}[n]\!=\!{\bf x}_k\big] 
	&\!\!= \! \frac{1}{(\pi \sigma^2)^{N_{\rm rx}}} \!\exp\!\left(\! -\frac{\|{\bf y}[n] \!-\! {\bf H}^{\sf H}{\bf x}_k\|^2}{\sigma^2}\right)\!,
%	&\!= \! \frac{\exp\!\left( -\frac{\|{\bf y}[n] - {\bf H}^{\sf H}{\bf x}_k\|^2}{\sigma^2}\right)}{(\pi \sigma^2)^{N_{\rm rx}}}   ,
\end{align}
for $k\in\mathcal{K}$. Here, we assume that the probability of transmitting each symbol vector is equal (i.e., $\mathbb{P}\big[{\bf x}[n]\!=\!{\bf x}_k\big]=\frac{1}{K}$, $\forall k\in\mathcal{K}$). As can be seen from \eqref{eq:APP_def} and \eqref{eq:LF_def}, if the true channel ${\bf H}$ is known at the receiver, the exact APPs are obtained from the MAP data detector. Unfortunately, in practical communication systems, the information of ${\bf H}$ is infeasible at the receiver due to channel estimation errors; these errors are inevitable when employing conventional pilot-aided channel estimators with a limited length of the pilot signals. Since the performance of the data detector heavily depends on the accuracy of the channel information at the receiver, developing a proper method to reduce the channel estimation error is essential to maximize the data detection performance. 
\section{Proposed LMMSE Channel Estimator}\label{Sec:Proposed}
In this section, we propose a data-aided LMMSE channel estimator that improves the MSE performance of the conventional LMMSE channel estimator based on a reinforcement learning approach.
%by exploiting detected symbol vectors obtained from the data detection as additional pilot signals. To maximize the performance of the proposed method, only the symbol vectors that are correctly detected at the receiver should be exploited as pilot signals, because the channel information obtained from incorrectly detected symbol vectors is . 

%%%%%%%%%%%%%%%%%%%%%%%%%%%%%%%%%%%%%%%%%%%%%%%%%%%%%%%%%%%%%%%%%%%%%%%%%%%%%%%%%%%%%%%%%%%%
%%%%%%%%%%%%%%%%%%%%%%%%%%%%%%%%%%%%%%%%%%%%%%%%%%%%%%%%%%%%%%%%%%%%%%%%%%%%%%%%%%%%%%%%%%%%
\subsection{Basic Idea}\label{Sec:LMMSE} 
The basic idea is to update the LMMSE channel estimate by selectively exploiting detected symbol vectors obtained from the data detection as additional pilot signals. For example, after the data detection of the $b$-th data block, the receiver obtains the set of detected symbol vectors $\{\hat{\bf x}[n]\}_{n\in\mathcal{N}_b}$ and the corresponding received signals $\{{\bf y}[n]\}_{n\in\mathcal{N}_b}$, where $\hat{\bf x}[n]$ is the detected symbol vector at time slot $n$ defined as
\begin{align}\label{eq:x_hat_def}
	\hat{\bf x}[n] = \argmax_{{\bf x}_k\in\mathcal{X}^{N_{\rm tx}}}~ \mathbb{P}\big[{\bf x}[n]\!=\!{\bf x}_k\big|{\bf y}[n]\big].
\end{align}
If all the detected symbol vectors are the same with the transmitted symbol vectors (i.e., $\hat{\bf x}[n]={\bf x}[n]$, $\forall n\in\mathcal{N}_b$), exploiting $\{\hat{\bf x}[n]\}_{n\in\mathcal{N}_b}$ as additional pilot signals gives the new LMMSE channel estimate:
\begin{align}
	\hat{\bf h}_{{\rm new},r} = \big({\bf X}_{\rm new} {\bf X}_{\rm new}^{\sf H} + \sigma^2 {\bf I}_{N_{\rm tx}}\big)^{-1}{\bf X}_{\rm new}\bar{\bf y}_{{\rm new},r}^{\sf H},
\end{align}
where $\bar{\bf y}_{{\rm new},r} = \big[{\bf y}_{r}^{\rm p},y_r[\mathcal{N}_b(1)],\cdots,y_r[\mathcal{N}_b(T_{\rm d})]\big]$, ${\bf X}_{\rm new} = \big[ {\bf X}^{\rm p}, \hat{\bf x}[\mathcal{N}_b(1)],\cdots,  \hat{\bf x}[\mathcal{N}_b(T_{\rm d})]\big]$,
%\begin{align}
%	\bar{\bf y}_{{\rm new},r} &= \big[{\bf y}_{r}^{\rm p},y_r[\mathcal{N}_b(1)],\cdots,y_r[\mathcal{N}_b(T_{\rm d})]\big],\\
%	{\bf X}_{\rm new} &= \big[ {\bf X}^{\rm p}, \hat{\bf x}[\mathcal{N}_b(1)],\cdots,  \hat{\bf x}[\mathcal{N}_b(T_{\rm d})]\big], 
%\end{align}
and $\mathcal{N}_b(i)$ is the $i$-th smallest element in $\mathcal{N}_b$. The above channel estimate is expected to be more accurate than the conventional LMMSE channel estimate in \eqref{eq:h_hat_conv} because a larger number of the pilot signals are used to obtain this new estimate.

Unfortunately, in practical communication systems, some detected symbol vectors may differ from the transmitted symbol vectors (i.e., $\hat{\bf x}[n]\neq{\bf x}[n]$) due to data detection errors. In addition, whether each symbol vector is correctly detected is generally unknown at the receiver. Exploiting such incorrect symbol vectors as additional pilot signals may degrade the accuracy of the channel estimate. Therefore, the major challenge of designing the data-aided channel estimator is to optimize the selection of the detected symbol vectors without knowing which vectors are correctly detected at the receiver.
%Unfortunately, in practical communication systems, some detected symbol vectors may differ from the transmitted symbol vectors (i.e., $\hat{\bf x}[n]\neq{\bf x}[n]$) due to data detection errors; thereby, exploiting these incorrect symbol vectors as additional pilot signals may degrade the accuracy of the channel estimate. In addition, whether each symbol vector is correctly detected is generally unknown at the receiver. Therefore, a major challenge underlying the above idea is to select a proper set of the detected symbol vectors that can be exploited as pilot signals without knowing which symbol vectors are correctly detected at the receiver.

%%%%%%%%%%%%%%%%%%%%%%%%%%%%%%%%%%%%%%%%%%%%%%%%%%%%%%%%%%%%%%%%%%%%%%%%%%%%%%%%%%%%%%%%%%%%
%%%%%%%%%%%%%%%%%%%%%%%%%%%%%%%%%%%%%%%%%%%%%%%%%%%%%%%%%%%%%%%%%%%%%%%%%%%%%%%%%%%%%%%%%%%%
\subsection{Optimization Problem: Markov Decision Process}\label{Sec:MDP} 
To deal with the aforementioned challenge, we formulate an optimization problem that finds the best selection of the detected symbol vectors to maximize the accuracy of the channel estimate when they are exploited as additional pilot signals. Particularly, we formulate this problem as a Markov decision process (MDP) to make a sequential decision on the use of each detected symbol vector while considering the effect of the error propagation caused by the current decision on the decisions for subsequent symbol vectors. %Each component of the MDP is defined below.

%%%%%%%%%%%%%%%%%%%%%%%%%%%%%%%%%%%%%%%%%%%%%%%%%%%%%%%%%%%%%%%%%%%%%%%%%%%%%%%%%%%%%%%%%%%%
\subsubsection{State}
The state set of the MDP associated with time slot $n$ is defined as 
\begin{align}\label{eq:State_set}
	\mathcal{S}_n
	\!=\! \big\{ ({\bf X}_n,\hat{\bf X}_n,\mathcal{M})~&\big|
	~{\bf X}_n \!=\!\big[{\bf X}^{\rm p}, {\bf x}_{k_1},\cdots,{\bf x}_{k_{|\mathcal{M}|}}\big], k_i \!\in\! \mathcal{K},   \nonumber \\
	&~~\hat{\bf X}_n \!=\!\big[{\bf X}^{\rm p}, \hat{\bf x}[\mathcal{M}(1)],\!\cdots\!,\hat{\bf x}[\mathcal{M}(|\mathcal{M}|)\big],\nonumber \\ 
	&~~ \mathcal{M} \!\subset\! \{T_{\rm p}\!+\!1,\ldots,n\!-\!1\} \big\},
\end{align}
where $\mathcal{M}(i)$ is the $i$-th smallest element in $\mathcal{M}$. In \eqref{eq:State_set}, the subset $\mathcal{M}$ represents the set of time slot indices associated with the detected symbol vectors that will be exploited as additional pilot signals while being transmitted before time slot $n$.
%In \eqref{eq:State_set}, ${\bf X}_n$ and $\hat{\bf X}_n$ represent the concatenated matrices that consist of the transmitted and the detected symbol vectors, respectively, which are exploited as additional pilot signals until time slot $n$. In addition, the subset $\mathcal{M}$ of $\{T_{\rm p}\!+\!1,\ldots,n\}$ in \eqref{eq:State_set} represents the set of time slot indices associated with the symbol vectors selected until time slot $n$. 
Using this definition, the LMMSE channel estimate obtained at the state ${\rm S}_n=({\bf X}_n,\hat{\bf X}_n,\mathcal{M})\in\mathcal{S}_n$ is given by 
\begin{align}\label{eq:y_update}
	\hat{\bf h}_r({\rm S}_n) = \big(\hat{\bf X}_n \hat{\bf X}_n^{\sf H} + \sigma^2 {\bf I}_{N_{\rm tx}}\big)^{-1}\hat{\bf X}_n\bar{\bf y}_r^{\sf H}({\rm S}_n),
\end{align}
where $\bar{\bf y}_r({\rm S}_n) = \big[{\bf y}_{r}^{\rm p},y_r[\mathcal{M}(1)],\cdots,y_r[\mathcal{M}(|\mathcal{M}|)]\big]$.
%\begin{align}
%%	{\bf W}({\rm S}_n) &= (\hat{\bf X}_n \hat{\bf X}_n^{\sf H} + \sigma^2 {\bf I}_{N_{\rm tx}})^{-1}\hat{\bf X}_n, \label{eq:W_def}\\
%	\bar{\bf y}_r({\rm S}_n) &= \big[{\bf y}_{r}^{\rm p},y_r[m_1],y_r[m_2],\cdots,y_r[m_{|\mathcal{M}|}]\big], \label{eq:y_def}
%\end{align}

%%%%%%%%%%%%%%%%%%%%%%%%%%%%%%%%%%%%%%%%%%%%%%%%%%%%%%%%%%%%%%%%%%%%%%%%%%%%%%%%%%%%%%%%%%%%
\subsubsection{Action}
The action set of the MDP is defined as $\mathcal{A} = \{1,0\}$ which indicates whether to exploit the current detected symbol vector as an additional pilot signal. For
example, the action $a=1\in\mathcal{A}$ at the state ${\rm S}_n \in\mathcal{S}_n$ implies that the $n$-th detected symbol vector $\hat{\bf x}[n]$ will be exploited as the pilot signal.
%
%\begin{figure*}
%	\centering 
%	\subfigure[Original MDP]
%	{\epsfig{file=Fig_Tree_Ori.eps, height=6cm}}
%	\qquad\qquad\qquad
%	\subfigure[Approximate MDP]
%	{\epsfig{file=Fig_Tree_Approx.eps, height=6cm}}
%	\caption{The search tree of the MDP defined in Section IV-A with the original transition function in \eqref{eq:Trans2} (Fig.~\ref{fig:Tree}(a)) and with the approximate transition function in \eqref{eq:T_hat} (Fig.~\ref{fig:Tree}(b)).} \vspace{-3mm}
%	\label{fig:Tree}
%\end{figure*}

%%%%%%%%%%%%%%%%%%%%%%%%%%%%%%%%%%%%%%%%%%%%%%%%%%%%%%%%%%%%%%%%%%%%%%%%%%%%%%%%%%%%%%%%%%%%
\subsubsection{Transition Function}
From the definitions of the state and the action, the state transition function of the MDP for $a\in\mathcal{A}$ and ${\rm S}_n\in\mathcal{S}_n$ is represented as
\begin{align}\label{eq:Trans}
	{\sf T}^{(a,j)}({\rm S}_n) &=\mathbb{P}\big[{\sf U}_{n+1}^{(a,j)}({\rm S}_n) \big| {\rm S}_n,a\big] \nonumber \\
	&= \begin{cases}
		\mathbb{I}[{\bf x}[n]={\bf x}_j], &j\!\in\!\mathcal{J}_a, a\!=\!1,  \\
		1, & j\!\in\!\mathcal{J}_a, a\!=\!0.
	\end{cases}
\end{align}
where $\mathcal{J}_0=\{0\}$, $\mathcal{J}_1=\{1,\ldots,K\}$, and ${\sf U}_{n+1}^{(a,j)}({\rm S}_n)\in\mathcal{S}_{n+1}$ is the state that can be transited from the state ${\rm S}_n=({\bf X}_n,\hat{\bf X}_n,\mathcal{M}) \in\mathcal{S}_{n}$ with the action $a\in\mathcal{A}$, given by
\begin{align}\label{eq:S_trans}
	&{\sf U}_{n+1}^{(a,j)}({\rm S}_n)  \nonumber \\
	&= \begin{cases}
	\big([{\bf X}_n,{\bf x}_j], \big[\hat{\bf X}_n,\hat{\bf x}[n]\big],\mathcal{M}\cup\{n\}\big), &\!\!\!j\!\in\!\mathcal{J}_a, a\!=\!1, \\
	\big({\bf X}_n, \hat{\bf X}_n,\mathcal{M}\big),&\!\!\! j\!\in\!\mathcal{J}_a, a\!=\!0.
	\end{cases}
\end{align}

%%%%%%%%%%%%%%%%%%%%%%%%%%%%%%%%%%%%%%%%%%%%%%%%%%%%%%%%%%%%%%%%%%%%%%%%%%%%%%%%%%%%%%%%%%%%
\subsubsection{Reward Function}
The reward function of the MDP is defined as the MSE improvement between the channel estimate at the current state and the channel estimate at the next state. The MSE of the channel estimate for the $r$-th receive antenna at the state ${\rm S}_n\in\mathcal{S}_n$ is expressed as 
\begin{align}\label{eq:MSE}
	{\sf MSE}_r({\rm S}_n) = \!\mathbb{E} \Big[ \big\|\hat{\bf h}_r({\rm S}_n) - {\bf h}_r\big\|^2\Big] =  {\sf Tr}\big[{\bf C}_{\rm e}({\rm S}_n)\big],
\end{align}
where ${\bf C}_{\rm e}({\rm S}_n) = \mathbb{E}\big[\big(\hat{\bf h}_r({\rm S}_n) \!-\! {\bf h}_r\big)\big(\hat{\bf h}_r({\rm S}_n)\! -\! {\bf h}_r\big)^{\!\sf H}\big]$ for any $r\in\{1,\ldots,N_{\rm rx}\}$, and the expectation is taken with respect to the channel and the noise distributions. Note that ${\bf C}_{\rm e}({\rm S}_n)$ does not depend on a receive antenna index because we assume that the channel and the noise distributions are equal across different receive antennas. Then the reward function associated with the state transition from ${\rm S}_n \in \mathcal{S}_n$ to ${\rm S}_{n+1} \in \mathcal{S}_{n+1}$ is given by 
\begin{align}\label{eq:Reward}
	{\sf R}({\rm S}_n,{\rm S}_{n+1}) 
	&={\sf Tr}\left[{\bf C}_{\rm e}({\rm S}_n) - {\bf C}_{\rm e}({\rm S}_{n+1})\right].
\end{align}
%where the last equality is obtained from ${\bf C}_{\rm e}({\rm S}_n)={\bf C}_{\rm e}_r({\rm S}_n)$ for $r\in\{1,\ldots,N_{\rm rx}\}$ which holds because we assume that the channel and the noise distributions are equal across different receive antennas.
%Since we assume that the channel and the noise distributions are equal across different receive antennas, the reward funciton in \eqref{eq:Reward0} can be simplified to 
%\begin{align}\label{eq:Reward}
%	{\sf R}({\rm S}_n,{\rm S}_{n+1}) 
%	&= N_{\rm rx} {\sf Tr}\left[{\bf C}_{\rm e}({\rm S}_n) - {\bf C}_{\rm e}({\rm S}_{n+1})\right],
%\end{align}
%by denoting ${\bf C}_{\rm e}({\rm S}_n)={\bf C}_{\rm e}_r({\rm S}_n)$ for $r\in\{1,\ldots,N_{\rm rx}\}$. 
%\begin{align}
%	{\sf E}_r({\rm S}_n) &= \!\mathbb{E} \!\left[ \left( {\bf W}({\rm S}_n)\bar{\bf y}_r^{\sf H}({\rm S}_n) \!-\! {\bf h}_r \right) 
%	\!\left( {\bf W}({\rm S}_n)\bar{\bf y}_r^{\sf H}({\rm S}_n) \!-\! {\bf h}_r \right)^{\sf H}\right]\!, \label{eq:Error_Cov} 
%\end{align}
%The above reward function quantifies the MSE improvement in the channel estimate associated with the state transition from ${\rm S}_n \in \mathcal{S}_n$ to ${\rm S}_{n+1} \in \mathcal{S}_{n+1}$. 

%%%%%%%%%%%%%%%%%%%%%%%%%%%%%%%%%%%%%%%%%%%%%%%%%%%%%%%%%%%%%%%%%%%%%%%%%%%%%%%%%%%%%%%%%%%%
\subsubsection{Optimal Policy}
the optimal policy of the MDP is defined as
\begin{align}\label{eq:Policy_opt0}
	\pi^\star ({\rm S}_n) &= \argmax_{a \in \mathcal{A}}~ {\sf Q}({\rm S}_n,a),
\end{align}
where ${\sf Q}({\rm S}_n,a)$ is the Q-value that represents the optimal sum of the rewards obtained after taking the action $a\in\mathcal{A}$ at the state ${\rm S}_n\in\mathcal{S}_n$. By the definition of the transition function in \eqref{eq:Trans}, the Q-value is expressed as
%\begin{align}\label{eq:Q_value}
%	&{\sf Q}({\rm S}_n,a)\nonumber \\
%	&= \!\sum_{j\in\mathcal{J}_a} {\sf T}^{(a,j)}({\rm S}_n)  
%	\Big\{ {\sf R}\big({\rm S}_n,{\sf U}_{n+1}^{(a,j)}({\rm S}_n)\big) + {\sf V}^\star\big({\sf U}_{n+1}^{(a,j)}({\rm S}_n) \big)\Big\}\!,
%\end{align}
\begin{align}\label{eq:Q_value}
	{\sf Q}({\rm S}_n,a)
	&= \!\sum_{j\in\mathcal{J}_a} {\sf T}^{(a,j)}({\rm S}_n)  \nonumber \\
	&~~~\times\!\Big\{ {\sf R}\big({\rm S}_n,{\sf U}_{n+1}^{(a,j)}({\rm S}_n)\big)  
	 + {\sf V}^\star\big({\sf U}_{n+1}^{(a,j)}({\rm S}_n) \big)\Big\}\!,
\end{align}
where ${\sf V}^{\star}({\rm S}_n)$ for $n\in\mathcal{N}_b$ is the optimal sum of the future rewards when optimally acting from the state ${\rm S}_n$ until it reaches to  the state associated with the next data block 
(i.e., ${\rm S}_{\mathcal{N}_{b+1}(1)} \in \mathcal{S}_{\mathcal{N}_{b+1}(1)}$).
%, and $\gamma \in [0,1]$ is a discounting factor that determines the value of the future rewards in the current decision. 
Note that a finite-horizon and undiscounted problem is considered in this work because the goal of our problem is to maximize the accuracy of the channel estimate when the decisions are made on all the detected symbol vectors in each data block.
%In this work, we particularly set $\zeta = 0$, which yields
%\begin{align}\label{eq:Q_value}
%	{\sf Q}({\rm S},a)
%	&= \sum_{j\in\mathcal{K}\cup \{0\}} {\sf T}_{j}({\rm S},a) {\sf R}\big({\rm S},{\sf U}_{j}^\prime({\rm S},a)\big).
%\end{align}
%The reason for the above setting is that the XXXXXXXXXXXXX  

The above MDP cannot be solved using dynamic programming in practical communication systems. The reason is that the transition function in \eqref{eq:Trans} is unknown at the receiver due to the lack of information of the transmitted symbol vectors. Furthermore, solving this MDP may require a prohibitive computational complexity because the number of the states exponentially increases with the number of the detected symbol vectors in each data block (i.e., $T_{\rm d}$). Therefore, in what follows, we design a computationally-efficient algorithm to solve the above MDP which is applicable when the information of the true transition function is unknown.

%%%%%%%%%%%%%%%%%%%%%%%%%%%%%%%%%%%%%%%%%%%%%%%%%%%%%%%%%%%%%%%%%%%%%%%%%%%%%%%%%%%%%%%%%%%%
%%%%%%%%%%%%%%%%%%%%%%%%%%%%%%%%%%%%%%%%%%%%%%%%%%%%%%%%%%%%%%%%%%%%%%%%%%%%%%%%%%%%%%%%%%%%
\subsection{Proposed Solution: A Reinforcement-Learning Approach}\label{Sec:RL} 
Reinforcement learning is a promising technique to solve an MDP with unknown or partial information on the model \cite{RL:Book}. Inspired by this, we present a computationally-efficient algorithm that approximately but efficiently solves the MDP defined in Section~\ref{Sec:MDP}. Our strategy is to approximate both the transition and the value functions by exploiting the APPs obtained from the data detection. First, motivated by the fact that $\theta_{j}[n]$ is the APP of the event $\{{\bf x}[n]={\bf x}_j\}$, we approximate the transition function in \eqref{eq:Trans} as     
\begin{align}\label{eq:T_approx}
	{\sf T}^{(a,j)}({\rm S}_n) &\approx \begin{cases}
	\theta_j[n], &j\!\in\!\mathcal{J}_a, a\!=\!1,  \\
	1, & j\!\in\!\mathcal{J}_a, a\!=\!0.
	\end{cases}
\end{align}
The promising feature of the above approximation is that it approaches to the true transition function as the data detection performance improves. We also approximate the value function in \eqref{eq:Q_value} by considering a virtual state that mimics the optimal future behavior from the state ${\sf U}_{n+1}^{(a,j)}({\rm S}_n)$. Let $\tilde{\bf x}[n]$ be the soft-decision symbol vector at time slot $n$, defined as
\begin{align}\label{eq:x_soft}
	\tilde{\bf x}[n] = \sum_{k=1}^{K} \theta_k[n] {\bf x}_k,
\end{align}
for $n\in\mathcal{N}_b$ and $b\in\{1,\ldots,N_{\rm B}\}$. Using the above notation, we define the virtual state associated with time slot $m\geq n+2$ as 
\begin{align}\label{eq:S_hat}
	\hat{\sf U}_{m}^{(a,j)}({\rm S}_n) =\big({\bf X}_m^{(a,j)},\hat{\bf X}_m^{(a)},\mathcal{M}_m^{(a)}\big),
\end{align}
where
\begin{align*}
	{\bf X}_m^{(a,j)} %= \big[{\bf X}_{n+1}^{(a,j)},\tilde{\bf x}[n+1],\cdots,\tilde{\bf x}[m-1]\big] \nonumber \\
	&\!=\!\begin{cases}
	\big[{\bf X}_{n},{\bf x}_j,\tilde{\bf x}[n\!+\!1],\cdots,\tilde{\bf x}[m\!-\!1]\big], &\!\!\! j\!\in\!\mathcal{J}_a,a\!=\!1, \\
	\big[{\bf X}_{n},\tilde{\bf x}[n\!+\!1],\cdots,\tilde{\bf x}[m\!-\!1]\big], &\!\!\! j\!\in\!\mathcal{J}_a,a\!=\!0, \\
	\end{cases}   \\
	\hat{\bf X}_m^{(a)} % = \big[\hat{\bf X}_{n+1}^{(a,j)},\tilde{\bf x}[n+1],\cdots,\tilde{\bf x}[m-1]\big] \nonumber \\
	&\!=\!\begin{cases}
	\big[\hat{\bf X}_{n},\hat{\bf x}[n],\tilde{\bf x}[n\!+\!1],\cdots,\tilde{\bf x}[m\!-\!1]\big], &\!\!\! a\!=\!1, \\
	\big[\hat{\bf X}_{n},\tilde{\bf x}[n\!+\!1],\cdots,\tilde{\bf x}[m\!-\!1]\big], &\!\!\! a\!=\!0, \\
	\end{cases}  \\
	\mathcal{M}_m^{(a)}&\!=\!\begin{cases}
	\mathcal{M}\cup \{n+1,\ldots,m\}, & a=1, \\
	\mathcal{M}\cup \{n+2,\ldots,m\}, & a=0, \\
	\end{cases} 
\end{align*}
provided that ${\rm S}_n =({\bf X}_{n},\hat{\bf X}_{n},\mathcal{M})$. The intuition behind the virtual state in \eqref{eq:S_hat} is as follows: Suppose that the state ${\rm S}_{m}^{\star} = \hat{\sf U}_{m}^{(a,j)}({\rm S}_n) \in\mathcal{S}_m$ for $m\geq n+2$ is observed by optimally acting from the state $\hat{\sf U}_{n+1}^{(a,j)}({\rm S}_n)$ until time slot $m$. If the APP associated with the detected symbol vector at time slot $m$ is close to one, the optimal action is likely to be $\pi^\star({\rm S}_{m}^{\star})=1$ since the current detected symbol vector is reliable; in this case, the optimal state-action-state pair is approximated by $\big({\rm S}_{m}^{\star},\pi^\star({\rm S}_{m}^{\star}),{\rm S}_{m+1}^{\star}\big) \approx \big({\rm S}_{m}^{\star},1,\hat{\rm S}_{m+1}^{(a,j)}({\rm S}_n)\big)$  
%\begin{align}
%	\big({\rm S}_{m}^{\star},\pi^\star({\rm S}_{m}^{\star}),{\rm S}_{m+1}^{\star}\big) &\approx \big({\rm S}_{m}^{\star},1,\hat{\rm S}_{m+1}^{(a,j)}({\rm S}_n)\big),
%\end{align}
because ${\bf x}[m]\approx \tilde{\bf x}[m]\approx \hat{\bf x}[m]$. 
%$\big(\hat{\sf U}_{m}^{(a,j)}({\rm S}_n),\pi^\star(\hat{\sf U}_{m}^{(a,j)}({\rm S}_n)),{\sf U}_{m+1}^{\star}\big)\approx \big(\hat{\sf U}_{m}^{(a,j)}({\rm S}_n),1,\hat{\sf U}_{m+1}^{(a,j)}({\rm S}_n)\big)$ gives almost the same effect with the true state-action-state transition because $\theta_{\hat{k}[m]}[m]\approx 1$ implies that ${\bf x}[m]\approx \tilde{\bf x}[m]\approx {\bf x}_{\hat{k}[m]}$. 
Similarly, if the APP is evenly distributed across all symbol vectors at time slot $m$ (i.e., $\theta_{j}[m]\approx \frac{1}{K}$ for $j\in\mathcal{K}$), the optimal action is likely to be $\pi^\star({\rm S}_{m}^{\star})=0$ since the current detected symbol vector is unreliable; in this case, the optimal state-action-state pair is approximated by $\big({\rm S}_{m}^{\star},\pi^\star({\rm S}_{m}^{\star}),{\sf U}_{m+1}^{\star}\big) \approx \big({\rm S}_{m}^{\star},1,\hat{\rm S}_{m+1}^{(a,j)}({\rm S}_n)\big)$
%\begin{align}
%	\big({\rm S}_{m}^{\star},\pi^\star({\rm S}_{m}^{\star}),{\sf U}_{m+1}^{\star}\big) &\approx \big({\rm S}_{m}^{\star},1,\hat{\rm S}_{m+1}^{(a,j)}({\rm S}_n)\big),
%\end{align}
because exploiting the zero vector $\tilde{\bf x}[m] \approx {\bf 0}_{N_{\rm tx}}$ as the additional pilot signal is equivalent to not exploiting the symbol vector at time slot $m$.
%the optimal action is likely to be $a=1$ , and this effect is properly captured by $\hat{\sf U}_{m+1}^{(a,j)}({\rm S}_n) \leftarrow \hat{\sf U}_{m}^{(a,j)}({\rm S}_n)$. Similarly, if the APP is evenly distributed across all symbol vectors at time slot $m$ (i.e., $\theta_{j}[m]\approx \frac{1}{K}$ for $j\in\mathcal{K}$), we have $\tilde{\bf x}[m] \approx {\bf 0}_{N_{\rm tx}}$; in this case, the optimal action is likely to be $a=0$, and this effect is also captured by $\hat{\sf U}_{m+1}^{(a,j)}({\rm S}_n) \leftarrow \hat{\sf U}_{m}^{(a,j)}({\rm S}_n)$ because using the additional zero vector has no effect on the update of the channel estimate. 
Motivated by the above facts, we model the optimal future behavior from ${\sf U}_{n+1}^{(a,j)}({\rm S}_n)$ by considering the following virtual episode: 
\begin{align}\label{eq:Virtual_epi}
	&\Big({\sf U}_{n+1}^{(a,j)}({\rm S}_n),\pi^\star\big({\sf U}_{n+1}^{(a,j)}({\rm S}_n)\big),{\rm S}_{n+2}^{\star},\pi^\star({\rm S}_{n+2}^{\star}),\ldots,{\rm S}_{N_d^{\star}+1}^{\star}\Big) \nonumber \\
	&\approx \Big({\sf U}_{n+1}^{(a,j)}({\rm S}_n) ,1,\hat{\sf U}_{n+2}^{(a,j)}({\rm S}_n),1,\ldots,\hat{\sf U}_{N_d^{\star}+1}^{(a,j)}({\rm S}_n)\Big).
\end{align}
%By utilizing the virtual state, we approximate the value function ${\sf V}^\star\big({\sf U}_{n+1}^{(a,j)}({\rm S}_n) \big)$ as the sum of the rewards obtained when following a virtual episode of $\big({\sf U}_{n+1}^{(a,j)}({\rm S}_n) ,1,\hat{\sf U}_{n+2}^{(a,j)}({\rm S}_n),1,\ldots,\hat{\sf U}_{N_d^{\star}+1}^{(a,j)}({\rm S}_n)\big)$. 
%\begin{align}\label{eq:Virtual_epi}
%	\Big({\sf U}_{n+1}^{(a,j)}({\rm S}_n) ,1,\hat{\sf U}_{n+2}^{(a,j)}({\rm S}_n),1,\ldots,\hat{\sf U}_{N_d^{\star}+1}^{(a,j)}({\rm S}_n)\Big).
%\end{align}
Then we approximate the value function ${\sf V}^\star\big({\sf U}_{n+1}^{(a,j)}({\rm S}_n) \big)$ as the sum of the rewards obtained when following the virtual episode in \eqref{eq:Virtual_epi}:
\begin{align}\label{eq:V_approx}
	{\sf V}^\star\big({\sf U}_{n+1}^{(a,j)}({\rm S}_n)\big) 
	&\approx {\sf R}\big({\sf U}_{n+1}^{(a,j)}({\rm S}_n),\hat{\sf U}_{n+2}^{(a,j)}({\rm S}_n)\big) \nonumber \\
		&~~~+\! \sum_{m = n+2}^{\mathcal{N}_b(T_{\rm d})}\! {\sf R}\big(\hat{\sf U}_{m}^{(a,j)}({\rm S}_n),\hat{\sf U}_{m+1}^{(a,j)}({\rm S}_n)\big).
\end{align}
%\textcolor[rgb]{1,0,0}{Our strategy to approximate the transition function and the value function is illustrated in Fig.~XX.}

Based on the above strategy, we characterize the optimal policy for each state in a closed-form expression, as given in the following theorem:
\begin{thm}\label{lem:MSE_pro}
	Under the assumptions of \eqref{eq:T_approx} and \eqref{eq:V_approx}, the optimal policy for the state ${\rm S}_n=({\bf X}_n,\hat{\bf X}_n,\mathcal{M})\in\mathcal{S}_n$ is 
	\begin{align}\label{eq:Policy_pro}
		{\pi}^\star ({\rm S}_n) \!=\! \mathbb{I}\!\left[\frac{\sigma^2(1+\alpha_n) + \sigma^4\|{\bf t}_n\|^2 + \|{\bf v}_n\|^2 } 
			{2\sigma^4\beta_n + \delta_n + \|{\bf e}_n \!-\! {\bf u}_n \!+\! {\bf v}_n \|^2 } \geq 1\right],
		%\mathbb{I}\!\left[ {\rm Tr} \!\left( \!{\bf C}_{\rm e}\Big(\hat{\sf U}_{N_{d}^\star+1}^{(0,0)}({\rm S}_{n})\Big) 
		%	\!-\! \sum_{j=1}^K \theta_j[n] {\bf C}_{\rm e}\Big(\hat{\sf U}_{N_{d}^\star+1}^{(1,j)}({\rm S}_n)\Big) \!\right)\!\right],
%		{\pi} ({\rm S}) \!=\! \mathbb{I}\!\left[ {\rm Tr} \!\left( \!\hat{\bf C}_{\rm e}_{N_d^\star+1}^{(0,0)}({\rm S}_{n}) 
%			\!-\! \sum_{j=1}^K \theta_j[n] {\bf C}_{\rm e}_{N_d^\star+1}^{(1,j)}({\rm S}_n) \!\right)\!\right],
	\end{align}
	where ${\bf t}_n = {\bf Q}_{n}\hat{\bf x}[n]$, 
	${\bf u}_n = {\bf D}_{n}^{\sf H} {\bf t}_n$, 
	${\bf e}_n = \hat{\bf x}[n]-\tilde{\bf x}[n]$,
	%${\bf X}_0 = {\sf X}_{N_d^\star+1}^{(0,0)}({\rm S}_n)$, $\hat{\bf X}_{0} = \hat{\sf X}_{N_d^\star+1}^{(0,0)}({\rm S}_n)$, 
	\begin{align*}
		{\bf Q}_{n} &= \Bigg( \hat{\bf X}_{n}\hat{\bf X}_{n}^{\sf H} +\sum_{m=n+1}^{\mathcal{N}_b(T_{\rm d})}\tilde{\bf x}[m]\tilde{\bf x}^{\sf H}[m] + \sigma^2 {\bf I}_{N_{\rm tx}} \Bigg)^{-1},  \\
		{\bf D}_{n} &= \hat{\bf X}_{n} \big( \hat{\bf X}_{n} - {\bf X}_{n}\big)^{\sf H} + \sigma^2 {\bf I}_{N_{\rm tx}},  \\
		{\bf v}_n &= (1+\alpha_n)\frac{{\bf D}_{n}^{\sf H} {\bf Q}_n {\bf t}_n}{\|{\bf t}_n\|^2},\\
%		\alpha_n &= \hat{\bf x}^{\sf H}[n]{\bf Q}_n\hat{\bf x}[n], \\
%		\beta_n &=(1+\alpha_n)\frac{{\bf t}_n^{\sf H} {\bf Q}_n {\bf t}_n}{\|{\bf t}_n\|^2}, \\
		\delta_n &= \sum_{j=1}^{K} \theta_j[n]\big\|\hat{\bf x}[n]\!-\!{\bf x}_j\big\|^2 \!-\! \big\|\hat{\bf x}[n]\!-\!\tilde{\bf x}[n]\big\|^2, 
	\end{align*}
	$\alpha_n = \hat{\bf x}^{\sf H}[n]{\bf Q}_n\hat{\bf x}[n]$, and $\beta_n =(1+\alpha_n)\frac{{\bf t}_n^{\sf H} {\bf Q}_n {\bf t}_n}{\|{\bf t}_n\|^2}$.
\end{thm}
\begin{IEEEproof}
	See Appendix A.
\end{IEEEproof}

%The common feature of the approximations adopted in Theorem~1 is that their tightness increases as the data detection performance improves because $\hat{\sf T}_{a}({\sf U}_{j}^\prime({\rm S}),{\rm S}) \rightarrow {\sf T}_{a}({\sf U}_{j}^\prime({\rm S}),{\rm S})$ and $\hat{\sf U}_{j,d}^\star({\rm S}) \rightarrow {\sf U}_{j,d}^\star({\rm S})$ as $\theta_{n,k[n]} \rightarrow 1$ for $n\in\mathcal{N}_b$. This implies that a mismatch in the policy caused by the use of the approximations in \eqref{eq:T_hat} and \eqref{eq:S_opt_hat} can be reduced by improving the accuracy of the likelihood function estimates. Fortunately, the accuracy of the proposed estimates is expected to increase as the number of the input-output samples increases. Therefore, the optimal policy in Theorem~1 becomes close to  the \textit{true} optimal policy as the presented algorithm is proceeded for multiple data blocks within the channel coherence time.
%

%%%%%%%%%%%%%%%%%%%%%%%%%%%%%%%%%%%%%%%%%%%%%%%%%%%%%%%%%%%%%%%%%%%%%%%%%%%%%%%%%%%%%%%%%%%%
%%%%%%%%%%%%%%%%%%%%%%%%%%%%%%%%%%%%%%%%%%%%%%%%%%%%%%%%%%%%%%%%%%%%%%%%%%%%%%%%%%%%%%%%%%%%
\subsection{Special Case: Symbol Vector Reconstruction}\label{Sec:CRC} 
In a special case when all information bits are correctly decoded at the channel decoder, the receiver is able to reconstruct all the transmitted symbol vectors by applying transmission procedures (e.g., channel encoding and symbol mapping) to the decoded information bits. Furthermore, the existence of the decoding error is readily checked by the CRC bits with high probability, as discussed in \cite{Jeon:arXiv:19}. Motivated by the above facts, when the CRC check is successful, we reconstruct the symbol vectors and then use all these vectors as additional pilot signals, instead of applying the reinforcement learning approach in Sec. III-C.

\subsection{Summary: Proposed Algorithm}\label{Sec:Alg}
In Algorithm~1, we summarize the proposed data-aided LMMSE channel estimator. 
\begin{algorithm}[h]
	\caption{The proposed LMMSE channel estimator.}\label{alg:Sum}
	{\small{\begin{algorithmic}[1]
		\STATE Set ${\bf H} \leftarrow \hat{\bf H}=\big[\hat{\bf h}_1,\cdots,\hat{\bf h}_{N_{\rm rx}}\big]$ from \eqref{eq:h_hat_conv}.
		\STATE Initialize ${\rm S}_1 = \big({\bf X}^{\rm p},{\bf X}^{\rm p},\emptyset\big)$. 
		\FOR {$b=1$ to $N_{\rm B}$}	
			%\STATE Update ${\bf H} \leftarrow \hat{\bf H}$.
			\STATE Compute $\theta_{j}[n]$ and $\hat{\bf x}[n]$ from \eqref{eq:APP_def} and \eqref{eq:x_hat_def}, $\forall n\in\mathcal{N}_b,j\in\mathcal{K}$.
			%\STATE Compute $\{\theta_{j}[n]\}_{n\in\mathcal{N}_b,j\in\mathcal{K}}$ and $\{\hat{\bf x}[n]\}_{n\in\mathcal{N}_b}$ from \eqref{eq:APP_def}. 
			\IF {the CRC check for the $b$-th data block is successful}
				\STATE Reconstruct the symbol vectors ${\bf x}_{\rm rec}[n]$, $\forall n\in\mathcal{N}_b$.
				\STATE Set ${\bf X}_{\rm rec}=\big[{\bf x}_{\rm rec}[\mathcal{N}_b(1)],\cdots,{\bf x}_{\rm rec}[\mathcal{N}_b(T_{\rm d})]\big]$. 
				\STATE Set ${\rm S}_{\mathcal{N}_{b+1}(1)} = \big([{\bf X},{\bf X}_{\rm rec}], [\hat{\bf X},{\bf X}_{\rm rec} ],\mathcal{M}\cup \mathcal{N}_b\big)$ provided that ${\rm S}_{\mathcal{N}_b(1)} = \big({\bf X}, \hat{\bf X},\mathcal{M}\big)$.
			\ELSE
				\FOR {$n\in \mathcal{N}_b$}
					\STATE Compute $a^\star = {\pi}^\star({\rm S}_n)$ from \eqref{eq:Policy_pro}.
					\STATE Set $j^\star = 0$ for $a^\star = 0$ and ${\bf x}_{j^\star} = \hat{\bf x}[n]$ for $a^\star = 1$.
					\STATE Update ${\rm S}_{n+1} \leftarrow {\sf U}_{n+1}^{(a^\star,j^\star)}({\rm S}_n)$ from \eqref{eq:S_trans}. 
				\ENDFOR
			\ENDIF
			\STATE Set ${\bf H} \leftarrow\hat{\bf H}=\big[\hat{\bf h}_1({\rm S}_n),\cdots,\hat{\bf h}_{N_{\rm rx}}({\rm S}_n)\big]$ from \eqref{eq:y_update}.
		\ENDFOR
	\end{algorithmic}}}
\end{algorithm}

\vspace{-2mm}
\noindent 
In Step 12 of Algorithm~1, when the optimal action is determined as $a^\star\!=\!1$, we consider the most-probable state transition that can be taken from the state ${\rm S}_n$, in order to model the true state transition which is unknown at the receiver. This modeling allows the receiver to follow the true state transition particularly when the data detection is sufficiently reliable.

\section{Simulation Results}\label{sec:simul}
In this section, using simulations, we evaluate the performance gain achieved by the proposed LMMSE channel estimator in Sec. III. Here, we consider a MIMO system operating with the MAP data detector in Sec. II-C when $(N_{\rm tx},N_{\rm rx},T_{\rm d},N_{\rm B})=(2,4,256,20)$. Particularly, 4-QAM is adopted for the symbol mapping, 16-bit CRC with the polynomial of $z^{16} + z^{15} + z^2 + 1$ is adopted for the CRC encoding/decoding, and the rate $\frac{1}{2}$ turbo code is adopted for the channel coding based on parallel concatenated codes with feedforward and feedback polynomial (15,13) in octal notation. We also consider a per-bit signal-to-noise ratio (SNR) defined as $E_b/N_0 = \frac{1}{\log_2 |\mathcal{X}|\sigma^2}$.

%For performance comparison, we consider three data detection methods: 1) the MAP detection explained in Section~\ref{Sec:MAP}, 2) the GAMP-based detection in \cite{Wen:16}, and 3) zero-forcing (ZF) detection. We refer to the MAP detection operating with the proposed likelihood function estimation method as \textit{robust ML}. 

\begin{figure}
	\centering
	\includegraphics[width=6cm]{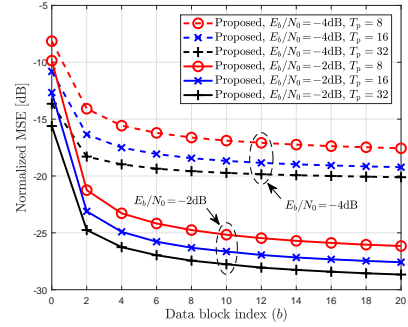} \vspace{-0.3cm}\caption{Normalized MSE vs. data block index of the proposed LMMSE channel estimator.} \vspace{-3mm} \label{fig:NMSE}
\end{figure}

Fig.~\ref{fig:NMSE} plots the normalized MSE (NMSE) of the channel estimate, computed as $\frac{\sum_{r=1}^{N_{\rm rx}} \|\hat{\bf h}_r - {\bf h}_r\|^2}{\sum_{r=1}^{N_{\rm rx}} \|{\bf h}_r\|^2}$,
%\begin{align}
%	\frac{\sum_{r=1}^{N_{\rm rx}} \|\hat{\bf h}_r - {\bf h}_r|^2}{\sum_{r=1}^{N_{\rm rx}} \|{\bf h}_r|^2}
%\end{align}
versus data block index $b$ of the proposed LMMSE channel estimator. Fig.~\ref{fig:NMSE} shows that the NMSE of the channel estimate significantly decreases with the data block index. It should be noticed that the NMSE at $b=0$ in Fig.~\ref{fig:NMSE} represents the estimation error of the conventional LMMSE channel estimator; thereby, the above result demonstrates that the estimation error reduction provided by the use of the proposed method increases as the number of detected symbol vectors at the receiver increases. It is also shown that a larger error reduction is achieved in the case of $E_b/N_0\!=\!-2$dB than in the case of $E_b/N_0=-4$ dB. The reason is that the number of reliable detected symbol vectors, that can be used as the additional pilot signals, increases as the data detection performance improves. 

%the amount of the error reduction in the case of $E_b/N_0\!=\!-2$dB (i.e., roughly $15$-dB reduction) is significantly less than that in the case of $E_b/N_0=-4$ dB (i.e., roughly $8$-dB reduction). The reason is that the number of reliable detected symbol vectors, that can be used as the aditional pilot signals, increases as the data detection performance improves. 

\begin{figure}
	\centering
	\includegraphics[width=6cm]{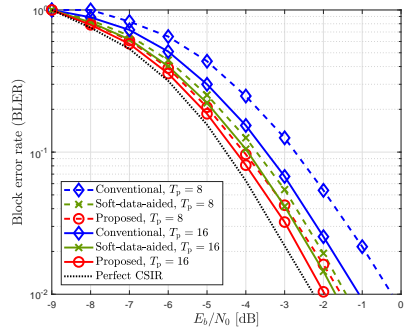} \vspace{-0.3cm}\caption{BLER vs. SNR of the conventional and the proposed LMMSE channel estimators.} \vspace{-3mm} \label{fig:BLER}
\end{figure}

Fig.~\ref{fig:BLER} compares the block-error-rates (BLERs) of the proposed and the conventional LMMSE channel estimators. For performance benchmark, the soft-data-aided LMMSE channel estimator is also plotted which exploits the soft-decision symbol vectors in \eqref{eq:x_soft} as additional pilot signals. Fig.~\ref{fig:BLER} shows the BLER of the proposed estimator is better than those of the conventional and the soft-data-aided estimators regardless of pilot lengths and per-bit SNRs. This result demonstrates the effectiveness of the proposed method which properly optimizes the selection of the detected symbol vectors via reinforcement learning. %It is also shown that even with the short pilot length $T_{\rm p}=16$, the performance of the proposed channel estimator is very close to the optimal performance achieved with perfect CSIR. %when employing the propose channel estimator, the use of the short pilot $T_{\rm p}=16$ is sufficient to perform close to the optimal performance achieved when perfect channel information is available at the receiver. 
Another interesting observation is that the proposed estimator with $T_{\rm p}=8$ even performs better than the conventional estimator with $T_{\rm p}=16$, which implies that the proposed channel estimator requires fewer pilot signals to achieve the same BLER performance.

\section{Conclusion}
In this paper, we have presented a data-aided LMMSE channel estimator for MIMO systems, which selectively exploits detected symbol vectors obtained from data detection as additional pilot signals. It has been shown that reinforcement learning provides an effective framework to optimize the selection of the detected symbol vectors, which allows the receiver to mitigate the error propagation effect without taking an iterative approach. Simulation results have demonstrated that the presented estimator significantly reduces the MSE of the channel estimate and therefore provides a better detection performance, compared to the conventional LMMSE channel estimator.

%The key finding is that temporal channel variations can be tracked by exploiting input-output samples obtained from the data detection. It is also shown that reinforcement learning provides an effective framework to optimize the exploitation of the input-output samples containing uncertainties. Simulation results have demonstrated that the presented method compensates for the effects of both the channel estimation error and the temporal channel variations.  %An important direction for future research is to extend the presented detector for the use in wideband MIMO systems that operate with low-precision ADCs beyond one-bit precision. %It would also be interesting to develop a low-complexity variation of the presented method for massive MIMO systems.

\appendices
%%%%%%%%%%%%%%%%%%%%%%%%%%%%%%%%%%%%%%%%%%%%%%%%%%%%%%%%%%%%%%%%%%%%%%%%%%%%%%%%%%%%%%%%%%%%
%%%%%%%%%%%%%%%%%%%%%%%%%%%%%%%%%%%%%%%%%%%%%%%%%%%%%%%%%%%%%%%%%%%%%%%%%%%%%%%%%%%%%%%%%%%%
%%%%%%%%%%%%%%%%%%%%%%%%%%%%%%%%%%%%%%%%%%%%%%%%%%%%%%%%%%%%%%%%%%%%%%%%%%%%%%%%%%%%%%%%%%%%
\section{Proof of Theorem~1}\label{Apdx:Thm1}
Suppose that ${\rm S}_n=\big({\bf X}_{n},\hat{\bf X}_{n},\mathcal{M}\big)$ and $n\in \mathcal{N}_b$. By applying \eqref{eq:Reward} into \eqref{eq:V_approx}, the value function at the state  ${\sf U}_{n+1}^{(a,j)}({\rm S}_n)\in\mathcal{S}_{n+1}$ is expressed as 
\begin{align}\label{eq:Apdx:V_star}
	\!\!{\sf V}^\star\big({\sf U}_{n+1}^{(a,j)}({\rm S}_n)\big)  
	&\!=\! {\rm Tr}\bigg[ {\bf C}_{\rm e}\big({\sf U}_{n+1}^{(a,j)}({\rm S}_n)\big) -  \hat{\bf C}_{\rm e}\big(\hat{\sf U}_{n+2}^{(a,j)}({\rm S}_n)\big) \nonumber \\
	&~~+\! \sum_{m = n+2}^{\mathcal{N}_b(T_{\rm d})}\!  {\bf C}_{\rm e}\big(\hat{\sf U}_{m}^{(a,j)}({\rm S}_n)\big) \!-\!  {\bf C}_{\rm e}\big(\hat{\sf U}_{m+1}^{(a,j)}({\rm S}_n)\big)\bigg]  \nonumber \\
	&\!=\! {\rm Tr}\Big[{\bf C}_{\rm e}\big({\sf U}_{n+1}^{(a,j)}({\rm S}_n)\big) \!-\!  {\bf C}_{\rm e}\big(\hat{\sf U}_{\mathcal{N}_b(T_{\rm d})+1}^{(a,j)}({\rm S}_n)\big)\Big].
\end{align}
For notational simplicity, we denote $\hat{\sf U}_{\mathcal{N}_b(T_{\rm d}) +1}^{(a,j)}({\rm S}_n)=\hat{\rm S}_{\rm end}^{(a,j)}=\big({\bf X}_{\rm end}^{(a,j)},\hat{\bf X}_{\rm end}^{(a)},\mathcal{M}_{\rm end}^{(a)}\big)$. Applying \eqref{eq:Reward}, \eqref{eq:T_approx}, and \eqref{eq:Apdx:V_star} into \eqref{eq:Q_value} yields 
\begin{align}\label{eq:Apdx:Q_value}
	{\sf Q}({\rm S}_n,a)
	&=  \sum_{j\in\mathcal{J}_a} \theta_j[n] 
	 {\rm Tr}\Big[ {\bf C}_{\rm e}\big({\rm S}_n\big) \!-\! {\bf C}_{\rm e}\big(\hat{\rm S}_{\rm end}^{(a,j)}\big) \Big].
\end{align}
Then the optimal policy in \eqref{eq:Policy_opt0} is expressed as
\begin{align}\label{eq:Apdx:Policy_opt}
	\pi^\star ({\rm S}_n) 
	&= \argmax_{a \in \{0,1\} }~ {\sf Q}({\rm S}_n,a)= \mathbb{I} \left[ {\sf Q}({\rm S}_n,1) \!-\! {\sf Q}({\rm S}_n,0) \geq 0\right] \nonumber \\
	&= \mathbb{I} \!\!\left[{\rm Tr}\Bigg[ {\bf C}_{\rm e}\big(\hat{\rm S}_{\rm end}^{(0,0)}\big)
		\!-\! \sum_{j=1}^{K} \theta_j[n] {\bf C}_{\rm e}\big(\hat{\rm S}_{\rm end}^{(1,j)}\big) \Bigg] \!\!\geq\! 0\right] \!\!.
\end{align}
As can be seen in the above, the optimal policy is determined by the difference between the expected MSEs with the action $a=0$ and the action $a=1$ at the ending state. 

From \eqref{eq:y_n}, the distribution of $\bar{\bf y}_r^{\sf H}\big( \hat{\rm S}_{\rm end}^{(a,j)}\big)$ is given by 
\begin{align*}
%	{\bf W}\big( \hat{\rm S}_{\rm end}^{(a,j)}\big) 
%	&= \Big(\hat{\bf X}_{\rm end}^{(a)}\big(\hat{\bf X}_{\rm end}^{(a)}\big)^{\!\sf H} \!+ \sigma^2{\bf I}_{N_{\rm rx}}\Big)^{\!-1}\hat{\bf X}_{\rm end}^{(a)}, \label{eq:W_aj}\\
	\bar{\bf y}_r^{\sf H}\big( \hat{\rm S}_{\rm end}^{(a,j)}\big) 
	&\sim \mathcal{CN}\Big( {\bf 0}_{|\mathcal{M}_{\rm end}^{(a)}|},\big({\bf X}_{\rm end}^{(a,j)}\big)^{\!\sf H}{\bf X}_{\rm end}^{(a,j)} + \sigma^2{\bf I}_{|\mathcal{M}_{\rm end}^{(a)}|}\Big), 
\end{align*}
for $j\in\mathcal{J}_a$ and $a\in\mathcal{A}$. Using this fact, each error covariance matrix in \eqref{eq:Apdx:Policy_opt} is computed as 
%To compute the error covariance matrices in \eqref{eq:Apdx:Policy_opt}, it should be first noticed from \eqref{eq:y_n} that the distribution of $\bar{\bf y}_r^{\sf H}\big( \hat{\rm S}_{\rm end}^{(a,j)}\big)$ is given by 
%\begin{align*}
%%	{\bf W}\big( \hat{\rm S}_{\rm end}^{(a,j)}\big) 
%%	&= \Big(\hat{\bf X}_{\rm end}^{(a)}\big(\hat{\bf X}_{\rm end}^{(a)}\big)^{\!\sf H} \!+ \sigma^2{\bf I}_{N_{\rm rx}}\Big)^{\!-1}\hat{\bf X}_{\rm end}^{(a)}, \label{eq:W_aj}\\
%	\bar{\bf y}_r^{\sf H}\big( \hat{\rm S}_{\rm end}^{(a,j)}\big) 
%	&\sim \mathcal{CN}\Big( {\bf 0}_{|\mathcal{M}_{\rm end}^{(a)}|},\big({\bf X}_{\rm end}^{(a,j)}\big)^{\!\sf H}{\bf X}_{\rm end}^{(a,j)} + \sigma^2{\bf I}_{|\mathcal{M}_{\rm end}^{(a)}|}\Big), 
%\end{align*}
%for $j\in\mathcal{J}_a$ and $a\in\mathcal{A}$. Utilizing the above fact, each error covariance matrix in \eqref{eq:Apdx:Policy_opt} is computed as 
\begin{align}\label{eq:Apdx:E_aj}
	{\bf C}_{\rm e}\big(\hat{\rm S}_{\rm end}^{(a,j)}\big) 
	&\!=\! \sigma^2 {\bf Q}_n^{(a)} \!-\! \sigma^4 \big({\bf Q}_n^{(a)}\big)^{\!2} \!+\! {\bf Q}_n^{(a)}{\bf D}_n^{(a,j)}\big({\bf D}_n^{(a,j)}\big)^{\!\sf H} {\bf Q}_n^{(a)}\!,
\end{align}
where 
\begin{align*}
	&{\bf Q}_{n}^{(a)} 
	= \Big( \hat{\bf X}_{\rm end}^{(a)}(\hat{\bf X}_{\rm end}^{(a)})^{\sf H} + \sigma^2 {\bf I}_{N_{\rm tx}} \Big)^{-1} \nonumber  \\
	&\overset{(a)}{=}\begin{cases}
		\Big( \hat{\bf X}_{n}\hat{\bf X}_{n}^{\sf H} +\sum_{m=n+1}^{\mathcal{N}_b(T_{\rm d})}\tilde{\bf x}[m]\tilde{\bf x}^{\sf H}[m] + \sigma^2 {\bf I}_{N_{\rm tx}} \Big)^{\!-1}\!, &\!\!\! a=0,\\
		\Big( \big({\bf Q}_{n}^{(0)}\big)^{-1} + \hat{\bf x}[n]\hat{\bf x}^{\sf H}[n]\Big)^{\!-1}\!, &\!\!\!a=1,\\
	\end{cases}  \nonumber \\
	&{\bf D}_{n}^{(a,j)}
	= \hat{\bf X}_{\rm end}^{(a,j)} \Big( \hat{\bf X}_{\rm end}^{(a,j)} - {\bf X}_{\rm end}^{(a,j)}\Big)^{\sf H} + \sigma^2{\bf I}_{N_{\rm tx}} \nonumber \\
	&~~~~~~~\overset{(b)}{=}\begin{cases}
		\hat{\bf X}_{n} \big( \hat{\bf X}_{n} - {\bf X}_{n}\big)^{\sf H} + \sigma^2 {\bf I}_{N_{\rm tx}} , &\!\!j\in\mathcal{J}_a, a=0,\\
		{\bf D}_{n}^{(0,0)}+\hat{\bf x}[n] (\hat{\bf x}[n]\!-\!{\bf x}_j)^{\sf H},&\!\!j\in\mathcal{J}_a,a=1.\\
	\end{cases}  
\end{align*}
Note that the equalities of (a) and (b) are directly obtained from \eqref{eq:S_hat}. By the  matrix inversion lemma, the matrix ${\bf Q}_{n}^{(1)}$ is rewritten as 
\begin{align}\label{eq:Apdx:Q_np2}
	{\bf Q}_{n}^{(1)} 
	&\!= {\bf Q}_{n}^{(0)} \!+\! \frac{1}{ 1+\hat{\bf x}^{\sf H}[n]{\bf Q}_{n}^{(0)} \hat{\bf x}[n]} {\bf Q}_{n}^{(0)} \hat{\bf x}[n]\hat{\bf x}^{\sf H}[n]{\bf Q}_{n}^{(0)}\!,
\end{align}
In addition, by the definition of ${\bf D}_{n}^{(a,j)}$, we have
\begin{align}\label{eq:Apdx:D_np2}
	&\sum_{j=1}^{K} \theta_j[n]{\bf D}_{n}^{(1,j)}({\bf D}_{n}^{(1,j)})^{\sf H} \nonumber \\
	&= \big({\bf D}_{n}^{(0,0)} \!+\! \hat{\bf d}_n\big)\big({\bf D}_{n}^{(0,0)} \!+\! \hat{\bf d}_n\big)^{\!\sf H} + \delta_n\hat{\bf x}[n]\hat{\bf x}^{\sf H}[n],
\end{align}
where $\hat{\bf d}_n =\hat{\bf x}[n] (\hat{\bf x}[n]\!-\!\tilde{\bf x}[n])^{\sf H}$, and
\begin{align}
	\delta_n &= \sum_{j=1}^{K} \theta_j[n]\big\|\hat{\bf x}[n]\!-\!{\bf x}_j\big\|^2 \!-\! \big\|\hat{\bf x}[n]\!-\!\tilde{\bf x}[n]\big\|^2. \nonumber 
\end{align}
By applying \eqref{eq:Apdx:E_aj}, \eqref{eq:Apdx:Q_np2}, and \eqref{eq:Apdx:D_np2} into \eqref{eq:Apdx:Policy_opt}, we obtain the result in \eqref{eq:Policy_pro} where ${\bf Q}_n = {\bf Q}_{n}^{(0)}$ and ${\bf D}_{n}={\bf D}_{n}^{(0,0)}$.


\begin{thebibliography}{1}	
	\bibitem{Foschini:96} G. J. Foschini, ``Layered space-time architecture for wireless communication in a fading environment when using multi-element antennas,''
	{\em Bell Labs Tech. J.,} vol. 1, no. 2, pp. 41--59, 1996.
	
	\bibitem{Telatar:99} I. E. Telatar, ``Capacity of multi-antenna Gaussian channels,'' 
	{\em Europ. Trans. Telecommun.,} vol. 10, pp. 585--595, Nov./Dec. 1999.
	
	\bibitem{Zheng:03} L. Zheng and D. N. C. Tse, ``Diversity and multiplexing: A fundamental tradeoff in multiple-antenna channels,''
	{\em IEEE Trans. Inf. Theory,} vol. 49, no. 5, pp. 1073--1096, May 2003.
	
	%\bibitem{Sun:}Sumei Sun, I. Wiemer, C.K. Ho, and T. T. Tjhung, ``Training sequence assisted channel estimation for MIMO OFDMTraining sequence assisted channel estimation for MIMO OFDM,''
	
	\bibitem{Biguesh:06} M. Biguesh and A. B. Gershman, ``Training-based MIMO channel estimation: A study of estimator tradeoffs and optimal training signals,''
	{\em IEEE Trans. Sig. Process.,} vol. 54, no. 3, pp. 884--893, Mar. 2006.
	
	\bibitem{Ozdemir:07} M. K. Ozdemir and H. Arslan, ``Channel estimation for wireless OFDM systems,''
	{\em IEEE Commun. Surv.\&Tut.,} vol 9, no. 2, pp. 18--48, 2Q 2007.
	
	
	\bibitem{Dowler:03} A. Dowler, A. Nix, and J. McGeehan, ``Data-derived iterative channel estimation with channel tracking for a mobile fourth generation wide area OFDM system,''
	in {\em Proc. IEEE Global Telecommun. Conf. (GLOBECOM),} Dec. 2003
	
	\bibitem{Zhao:08} M. Zhao, Z. Shi, and M. C. Reed, ``Iterative turbo channel estimation for OFDM system over rapid dispersive fading channel,''
	{\em IEEE Trans. Wireless Commun.,} vol. 7, no. 8, Aug. 2008.
	
	\bibitem{Ma:14} J. Ma and L. Ping, ``Data-aided channel estimation in large antenna systems,''
	{\em IEEE Trans. Sig. Process.,} vol. 62, no. 12, pp. 3111--3124, Jun. 2014.

	
	\bibitem{Park:15} S. Park, B. Shim, and J. W. Choi, ``Iterative channel estimation using virtual pilot signals for MIMO-OFDM systems,''
	{\em IEEE Trans. Sig. Process.,} vol. 63, no. 12, pp. 3032--3045, Jun. 2015.
	
	\bibitem{Jeon:arXiv:19} Y.-S. Jeon, N. Lee, and H. V. Poor, ``Robust data detection for MIMO systems with one-bit ADCs: A reinforcement learning approach,''
	to be appeared in {\em IEEE Trans. Wireless Commun.,} 2020. 
	
	%%%%%%%%%%%%%%% In the main part
	\bibitem{RL:Book} R. S. Sutton and A. G. Barto, {\em Reinforcement Learning: An Introduction,} Cambridge, MA: The MIT Press, 2018.
	
%	\bibitem{Dong:04} M. Dong, L. Tong, and B. M. Sadler, ``Optimal insertion of pilot symbols for transmissions over time-varying flat fading channels,''
%	{\em IEEE Trans. Signal Process.,} vol. 52, no. 5, pp. 1403--1418, May 2004.

	%\bibitem{Pearl:98} J. Pearl, {\em Probabilistic Reasoning in Intelligent Systems: Networks of Plausible Inference,} Morgan Kaufmann, San Mateo, CA, 1988.  % No Cycle
	%
	%
	%\bibitem{IEEE} {\em IEEE Approved Draft Standard for LAN - Specific Requirements - Part II: Wireless LAN Medium Access Control (MAC) and Physical Layer (PHY) Specifications - Amendment 3: Enhancements for Very High Throughput in the 60GHz Band,} IEEE P802.11ad/D9.0 Std., Jul. 2012.
	%
	%%\bibitem{Rao} K. D. Rao, {\em Channel coding techniques for wireless communications,} Springer, 2015.
	%
	%\bibitem{Richardson} T. Richardson and R. Urbanke, {\em Modern coding theory,} Cambridge Univ. press, 2008.
	

\end{thebibliography}
\end{document}